%% file: paper.tex
\documentclass{PoS}
\usepackage[dvips]{graphicx}
\PoS{PoS(LAT2005)209}

\title{Heavy quark masses from Fermilab Fermions }

\ShortTitle{Heavy quark masses from Fermilab Fermions }

\author{\speaker{Matthew Nobes}\\
  Cornell University\\
  E-mail: \email{nobes@lepp.cornell.edu}}

\author{Howard Trottier\\
  Simon Fraser University\\
  E-mail: \email{trottier@sfu.ca}}

\abstract{
  Using automated perturbation theory techniques, we have computed the
  one-loop mass of Fermilab fermions, with an improved gluon
  action. We will present the results of these calculations, and the
  resulting predictions for the charm and bottom quark masses in the
  MSbar scheme.  We report $\overline{m}_{c}(\overline{m}_{c}) =
  1.22(9)$ GeV and $\overline{m}_{b}(\overline{m}_{b}) = 4.7(4)$ GeV.
  In addition we present results for the one-loop
  coefficients of the Fermilab action.
}

\FullConference{XXIIIrd International Symposium on Lattice Field Theory\\
  25-30 July 2005\\
  Trinity College, Dublin, Ireland}

\begin{document}

    In recent years there has been a lot of progress in lattice
    perturbation theory \cite{Trottier:2003bw,MasonPlen}.  In
    particular, 
    the implementation of
    automated perturbation theory routines has allowed many problems to be
    tackled.  Lattice perturbation theory has many applications, for example
    the determinations of improvement coefficients and renormalization
    factors.  In addition, lattice perturbation theory is used in
    connecting non-perturbative simulations to quantities of interest to
    the wider high-energy community.  An example of this is the recent
    determination of the strong coupling constant $\alpha_{s}$
    \cite{Mason:2005zx}. 

    In this report we present results of our one-loop calculations of
    current renormalizations, action parameters and quark masses for the
    Fermilab fermion action \cite{El-Khadra:1996mp}.  The lattice
    perturbation theory 
    produces an estimate of the quark's pole mass.  This can be
    converted into the $\overline{MS}$ scheme at some scale $\mu$.  We
    give a few details of this conversion and report preliminary values
    for the quark masses.  These are compared to the PDG values.

    Simulating heavy quarks on the lattice poses a special problem
    because the scale $m_{0}a$ is not small for typical lattice
    spacings.  For actions designed for light quarks
    $\mathcal{O}(m_{0}a)$ errors can be large.  There are a few
    different approaches to this problem.  One popular one is the
    Fermilab approach, which resums all mass dependance into the
    coefficients in the action.

    The Fermilab action is
    \begin{eqnarray}
      S & = & a^{4} \sum_{x} \bar{\psi}(x) \left[
	m_{0} + \frac{1+\gamma_{4}}{2} D_{4}^{-} -
	\frac{1-\gamma_{4}}{2} D_{4}^{+} +
	\zeta \vec{\gamma}\cdot\vec{D}  \right. \nonumber \\
	& - & \left. \frac{a r_{s} \zeta}{2} \triangle^{(3)} -
	\frac{iac_{B}\zeta}{2} \vec{\Sigma}\cdot\vec{B} -
	\frac{ac_{E}\zeta}{2} \vec{\alpha}\cdot\vec{E} \right] \psi(x)
    \end{eqnarray}
    where definitions of the various operators can be found in
    \cite{El-Khadra:1996mp}. 
    This action is designed to smoothly interpolate between the zero
    mass and infinite-mass limits.  As such it is ideal for simulations
    of charm quarks, where other methods (such as NRQCD) might have
    problems \cite{Trottier:1996ce}.  All the coefficients in the
    Fermilab action are mass 
    dependent, however at tree level we can use
    \begin{equation}
      \zeta = r_{s} = c_{E} = c_{B} = 1.
    \end{equation}
    
    Using our automated perturbation theory techniques we have completed
    all the ``basic'' one-loop calculations for this action.  These are
    the mass and wavefunction renormalizations, the renormalization of
    heavy-light and heavy-heavy vector and axial vector currents and the
    one-loop matching of the action parameters $c_{B}$ and $c_{E}$.  In
    the following sections we will present results for the action
    parameters and the mass renormalizations. 

    All of the calculations presented here were carried out with using
    our automated perturbation theory codes.  Apart from the
    wavefunction renormalization all these quantities are infrared
    finite and gauge invariant.  Infrared divergences in individual
    diagrams were regulated by using a gluon mass.

\begin{figure}
      \begin{center}
	\includegraphics{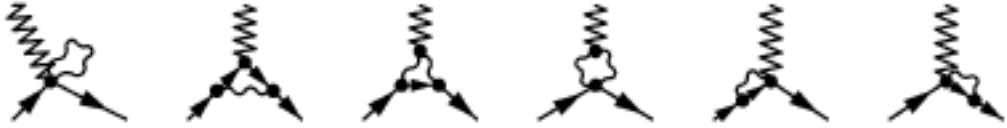}
      \end{center}
      \caption{Diagrams for matching action parameters}\label{vertmatch} 
    \end{figure}

    To match the coefficients $c_{B}(c_{E})$ we compute the scattering
    of a quark off of a background chromo-magentic(electric) field in
    both the lattice and continuum field theories, then tune the
    action parameters until the difference vanishes.  The relevant
    diagrams are shown in figure \ref{vertmatch}.

\begin{figure}
      \begin{center}
	\includegraphics[scale=0.65]{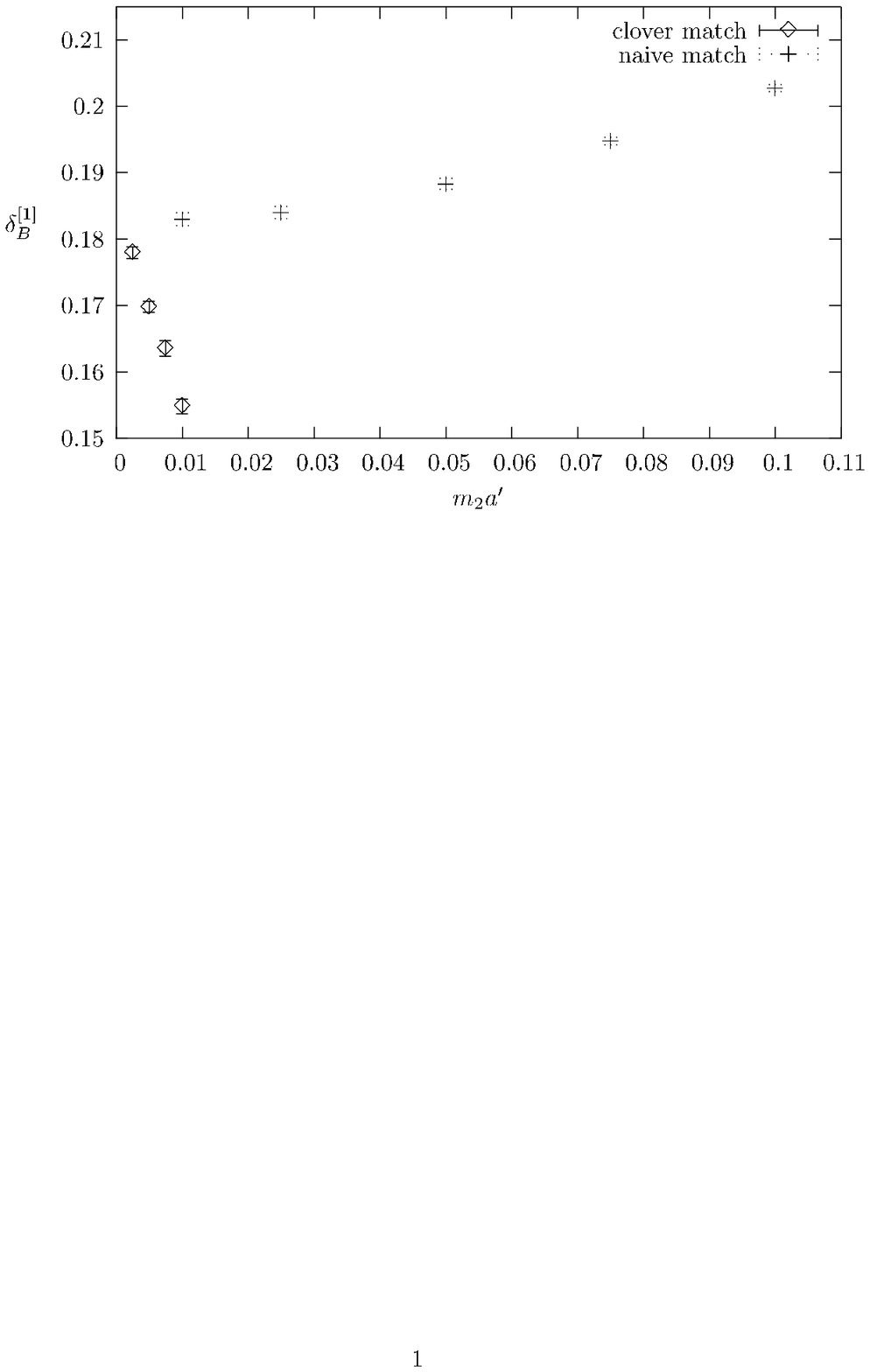}
      \end{center}
      \caption{One loop lattice to lattice matching of
	$\delta_{B} = c_{B} - \rm{counterterms}$ at $m_{0}a = 0.1$, 
	with improved glue.}\label{l2l}
    \end{figure}

    One interesting feature of our calculation is the use of lattice
    to lattice matching \cite{Nobes:2003nc}.  Rather than computing
    the continuum 
    contribution using standard methods, we use a simple lattice
    theory, with a spacing $a'$ that is driven very small.  Figure
    \ref{l2l} illustrates this for two cases, naive fermions (which
    have a quadratic approach to $a' = 0$) and Wilson fermions (which
    have a linear $a \to 0$ behaviour).  What is shown in figure \ref{l2l}
    is actually the difference between the two sets of diagrams
    (Fermilab at spacing $a$ and Wilson/naive at spacing $a'$) there
    are additional counterterms in the matching coming from the one
    loop part of $\zeta$, \cite{nobesphd} has details.  
    One sees that the same result for the matching coefficient $\delta_B$ is 
    obtained using the continuum limit ($a'\to0$) of either Wilson or naive quarks 
    for the continuum side of the matching.

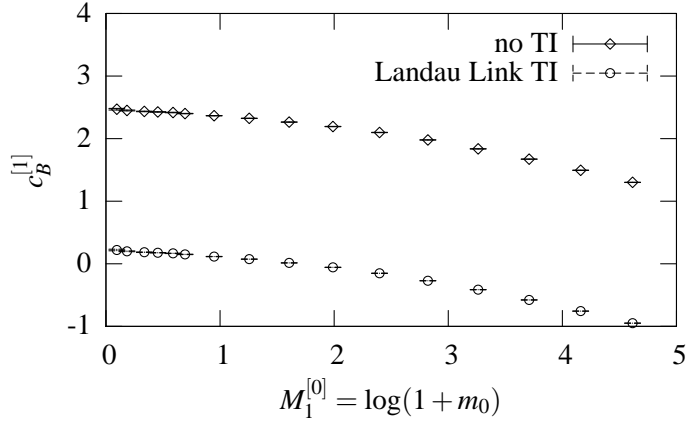
\begin{figure}
      \begin{center}
	\input{cb}
      \end{center}
      \caption{The one-loop coefficient of $\Sigma \cdot B$, with and
      without tadpole improvement}\label{cbol}
    \end{figure}

    The one-loop contribution to $c_{B}$ is plotted in figure
    \ref{cbol}.  The results for $c_{E}$ are very similar.  It is
    clear from the figure that the result, when 
    tadpole improved, is nearly zero over the whole range of
    interesting masses ($0<m_{0}a<2$).  This means that errors due to
    using only the tree level action parameters have likely been
    overestimated.  This conclusion only applies if the action has
    been tadpole improved.  The unimproved coefficients are quite
    large.

    To date all lattice determinations of the hyperfine splittings in
    the $J/\phi$ system have come out too low (\cite{Stewart:2000ev}
    and \cite{GottliebPost}).  These
    splittings are quite sensitive to the coefficient of $\Sigma\cdot
    B$ so it was believed that the one-loop determination of $c_{B}$
    would bring the splittings up.  This is not the case, the one-loop
    coefficient is very small.  However, there is evidence \cite{GottliebPost}
    that the discrepancy in the hyperfine splittings is decreasing as
    $a \to 0$.  A determination with the fully $\mathcal{O}(a^{2})$
    improved Fermilab action \cite{Oktay:2003gk} would be very useful.

    In addition to the action parameters, we have computed the quark
    masses for the Fermilab action.  This calculation is similar to
    \cite{Mertens:1997wx}, however we have used the Symanzik improved
    gluon action. 
    There are two masses to compute in the Fermilab formalism, the
    rest mass $M_{1}$ and the kinetic mass $M_{2}$.  These are defined
    via the small $pa$ expansion of the quark energy
    \begin{equation}
      E = M_{1} + \frac{\bf{p}^{2}}{2M_{2}} + \mathcal{O}(p^{4}a^{4}).
    \end{equation}

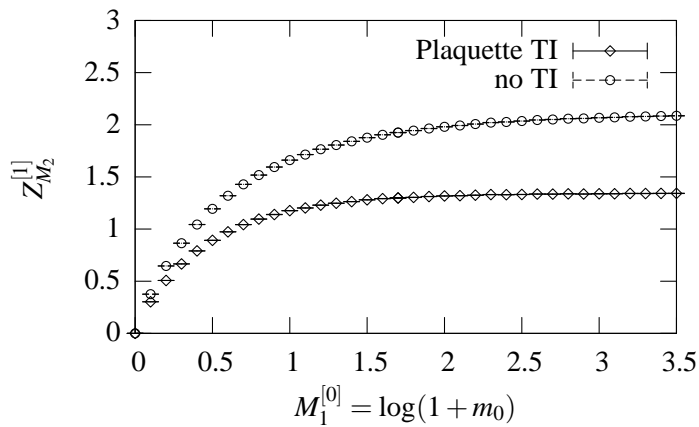
\begin{figure}
      \begin{center}
	\input{m11}
      \end{center}
      \caption{The one-loop rest mass, with and without tadpole improvement}\label{m11}
    \end{figure}

\begin{figure}
      \begin{center}
	\input{zm21}
      \end{center}
      \caption{The one-loop kinetic mass renormalization}\label{zm21}
    \end{figure}
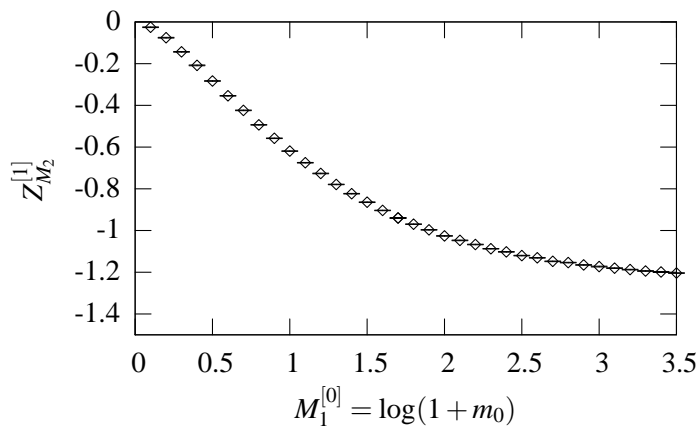

    The rest mass has the perturbative expansion
    $M_{1} = M_{1}^{[0]} + \alpha_{V}(q^{*}) M_{1}^{[1]} + \cdots$
    where $M_{1}^{[0]} = \log(1 + m_{0}a)$ and $\alpha_{V}(q^{*})$ is
    the QCD coupling in the V scheme evaluated at the BLM \cite{Hornbostel:2002af} 
    scale.  The
    kinetic mass is 
    usually expressed as follows
    \begin{equation}
      M_{2} = Z_{M_{2}} \frac{\exp{M_{1}}\sinh M_{1}}{1 + \sinh M_{1}}
    \end{equation}
    where $M_{1}$ is the \emph{all orders} rest mass and 
    \begin{equation}
      Z_{M_{2}} = 1 + \alpha_{V}(q^{*}) Z_{M_{2}}^{[1]} + \cdots .
    \end{equation}
    Figures \ref{m11} and \ref{zm21} show the one loop coefficients of
    the rest mass and the kinetic mass factor $Z_{M_{2}}$ over a wide
    range of input bare masses.  In all cases the we see a smooth
    transition from the small to large mass limits.
    

    The values $M_{1}$ and $M_{2}$ can be used to provide two
    different estimates of the pole mass of the quark
    \cite{Kronfeld:1997zc}.  The first 
    method is to estimate the binding energy, 
    $B_{1}a = M_{1}^{\bar{Q}Q',\rm{latt}} - N_{Q} M_{1}a$,
    where $M_{1}^{\bar{Q}Q',\rm{latt}}$ is a spin average meson mass
    computed on the lattice ($J/\psi/\eta_{c}$ for the $c$ quark,
    $B_{s}/B_{s}^{*}$ for the $b$ quark) and $N_{Q}$ is the number of
    heavy quarks in the meson (2 and 1, respectively).  The pole mass
    is then $m_{\rm{pole}} = M^{\bar{Q}Q',\rm{expt}} - B_{1}$,
    where $M^{\bar{Q}Q',\rm{expt}}$ is the meson mass taken from
    experiment.  Beyond the truncation of the perturbation series this
    method suffers from two major sources of error.  The first is that
    one needs to divide out the lattice spacing $a$, and the second is
    that it is quite sensitive to the bare mass $m_{0}$ used as input.

    The second method for estimating the pole mass does not suffer
    from these errors.  One begins with a perturbative determination
    of $M_{2}a$ and takes
    \begin{equation}
      m_{\rm{pole}} = \left(M_{2}a\right)^{\rm{PT}} 
      \frac{M^{\bar{Q}Q',\rm{expt}}}{M^{\bar{Q}Q',\rm{latt}}a}.
    \end{equation}
    This method avoids the errors of the method one, the dependance on
    the lattice spacing cancels and a small mistuning of the bare mass
    largely cancels in the ratio.  

    Once we have a value for the pole mass we can convert it into a
    value for the $\overline{MS}$ mass at some scale $\mu$ using
    \begin{equation}
      m_{\rm{pole}} = \overline{m}(\mu) \left\lbrace 
      1  + \frac{4}{3\pi} \left[ 
	1 + \log\left(
	\frac{\mu^{2}}{m_{0}^{2}}
	\right)
	\right]\alpha_{V}(q^{*})
      \right\rbrace
    \end{equation}
    We follow the conventional practice, and quote $\overline{m}(\mu)$
    at the scale of the $\overline{MS}$ mass itself, $\mu =
    \overline{m}$.  We use the BLM method to determine the scale
    $q^{*}$ in the coupling $\alpha_{V}(q^{*})$.  We estimate the
    relative systematic error coming from the neglected higher-orders
    in the perturbative matching as $\pm \alpha_{V}^{2}(q^{*})$.  We
    obtain
    \begin{eqnarray}
      \overline{m}^{\rm{method 1}}_{c}(\overline{m}_{c}) & = &
      1.24(1)(9) \, \rm{GeV} \quad \rm{Fine}\\ 
      \overline{m}^{\rm{method 2}}_{c}(\overline{m}_{c}) & = & 
      1.22(0)(9) \, \rm{GeV} \quad \rm{Fine}    \\ 
      \overline{m}^{\rm{method 1}}_{c}(\overline{m}_{c}) & = &
      1.39(1)(13) \, \rm{GeV} \quad \rm{Coarse}\\ 
      \overline{m}^{\rm{method 2}}_{c}(\overline{m}_{c}) & = &
      1.30(0)(12) \, \rm{GeV} \quad \rm{Coarse}\\
      \overline{m}^{\rm{method 1}}_{b}(\overline{m}_{b}) & = &
      4.4(1)(3) \, \rm{GeV} \quad \rm{Fine}\\ 
      \overline{m}^{\rm{method 2}}_{b}(\overline{m}_{b}) & = & 
      4.7(0)(4) \, \rm{GeV}\quad \rm{Fine}
    \end{eqnarray}
    where the first error is from the determination of the lattice
    spacing (which only affects method one) and the second our
    estimate of the unknown two-loop error.

    Our best values are the method two determinations on the fine
    lattice, which compare well with the PDG values
    $\overline{m}^{\rm{PDG}}_{c}(\overline{m}_{c}) = 
      1.25(10)$ GeV and
      $\overline{m}^{\rm{PDG}}_{b}(\overline{m}_{b}) = 
      4.25(15)$ GeV.
    These results are based on preliminary values for the input masses
    \cite{Simone:2005pc} so they may change somewhat.

    In this report we have presented results for the action parameters
    and masses of Fermilab fermions to one-loop.  For truly high
    precision determinations two-loop precision will be needed.  These
    calculations are in progress.

    \bibliography{refs} \bibliographystyle{JHEP-2}
\end{document}

%% file: cb.tex
%GNUPLOT: LaTeX picture with Postscript
\begin{picture}(0,0)%
\includegraphics{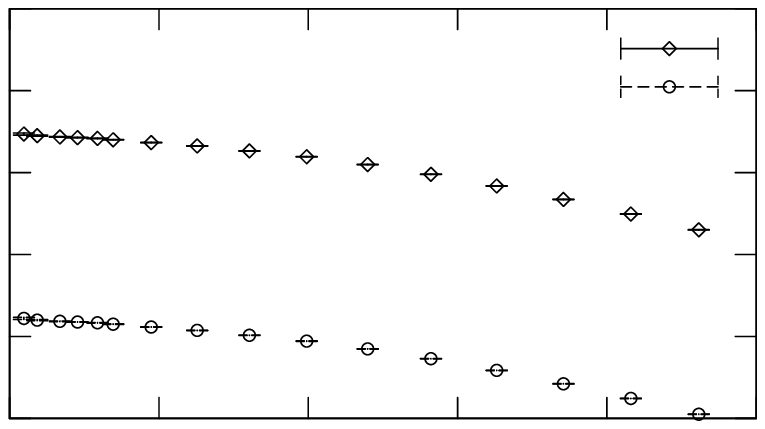}%
\end{picture}%
\begingroup
\setlength{\unitlength}{0.0200bp}%
\begin{picture}(13500,8100)(0,0)%
\put(1650,1650){\makebox(0,0)[r]{\strut{}-1}}%
\put(1650,2830){\makebox(0,0)[r]{\strut{} 0}}%
\put(1650,4010){\makebox(0,0)[r]{\strut{} 1}}%
\put(1650,5190){\makebox(0,0)[r]{\strut{} 2}}%
\put(1650,6370){\makebox(0,0)[r]{\strut{} 3}}%
\put(1650,7550){\makebox(0,0)[r]{\strut{} 4}}%
\put(1925,1100){\makebox(0,0){\strut{} 0}}%
\put(4075,1100){\makebox(0,0){\strut{} 1}}%
\put(6225,1100){\makebox(0,0){\strut{} 2}}%
\put(8375,1100){\makebox(0,0){\strut{} 3}}%
\put(10525,1100){\makebox(0,0){\strut{} 4}}%
\put(12675,1100){\makebox(0,0){\strut{} 5}}%
\put(550,4600){\rotatebox{90}{\makebox(0,0){\strut{}$c_{B}^{[1]}$}}}%
\put(7300,275){\makebox(0,0){\strut{}$M_{1}^{[0]} = \log(1 + m_{0})$}}%
\put(10450,6975){\makebox(0,0)[r]{\strut{}no TI}}%
\put(10450,6425){\makebox(0,0)[r]{\strut{}Landau Link TI}}%
\end{picture}%
\endgroup
 

%% file: m11.tex
%GNUPLOT: LaTeX picture with Postscript
\begin{picture}(0,0)%
\includegraphics{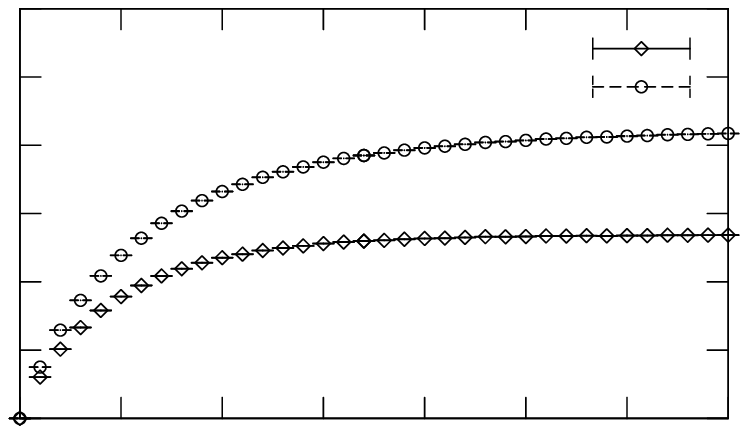}%
\end{picture}%
\begingroup
\setlength{\unitlength}{0.0200bp}%
\begin{picture}(13500,8100)(0,0)%
\put(2200,1650){\makebox(0,0)[r]{\strut{} 0}}%
\put(2200,2633){\makebox(0,0)[r]{\strut{} 0.5}}%
\put(2200,3617){\makebox(0,0)[r]{\strut{} 1}}%
\put(2200,4600){\makebox(0,0)[r]{\strut{} 1.5}}%
\put(2200,5583){\makebox(0,0)[r]{\strut{} 2}}%
\put(2200,6567){\makebox(0,0)[r]{\strut{} 2.5}}%
\put(2200,7550){\makebox(0,0)[r]{\strut{} 3}}%
\put(2475,1100){\makebox(0,0){\strut{} 0}}%
\put(3932,1100){\makebox(0,0){\strut{} 0.5}}%
\put(5389,1100){\makebox(0,0){\strut{} 1}}%
\put(6846,1100){\makebox(0,0){\strut{} 1.5}}%
\put(8304,1100){\makebox(0,0){\strut{} 2}}%
\put(9761,1100){\makebox(0,0){\strut{} 2.5}}%
\put(11218,1100){\makebox(0,0){\strut{} 3}}%
\put(12675,1100){\makebox(0,0){\strut{} 3.5}}%
\put(550,4600){\rotatebox{90}{\makebox(0,0){\strut{}$Z_{M_{2}}^{[1]}$}}}%
\put(7575,275){\makebox(0,0){\strut{}$M_{1}^{[0]} = \log(1 + m_{0})$}}%
\put(10450,6975){\makebox(0,0)[r]{\strut{}Plaquette TI}}%
\put(10450,6425){\makebox(0,0)[r]{\strut{}no TI}}%
\end{picture}%
\endgroup
 

%% file: zm21.tex
%GNUPLOT: LaTeX picture with Postscript
\begin{picture}(0,0)%
\includegraphics{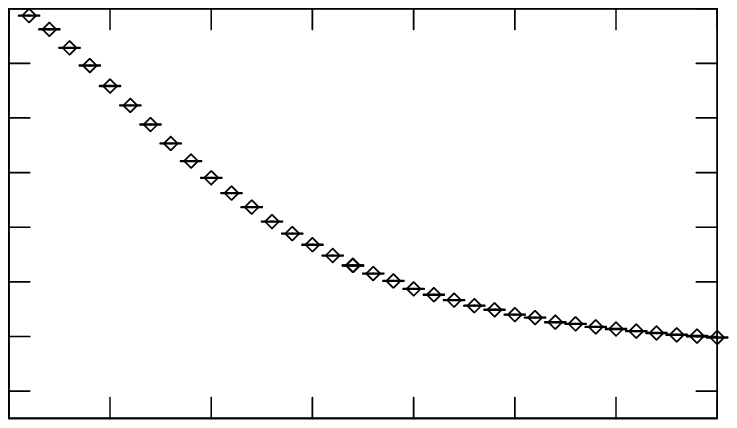}%
\end{picture}%
\begingroup
\setlength{\unitlength}{0.0200bp}%
\begin{picture}(13500,8100)(0,0)%
\put(2200,2043){\makebox(0,0)[r]{\strut{}-1.4}}%
\put(2200,2830){\makebox(0,0)[r]{\strut{}-1.2}}%
\put(2200,3617){\makebox(0,0)[r]{\strut{}-1}}%
\put(2200,4403){\makebox(0,0)[r]{\strut{}-0.8}}%
\put(2200,5190){\makebox(0,0)[r]{\strut{}-0.6}}%
\put(2200,5977){\makebox(0,0)[r]{\strut{}-0.4}}%
\put(2200,6763){\makebox(0,0)[r]{\strut{}-0.2}}%
\put(2200,7550){\makebox(0,0)[r]{\strut{} 0}}%
\put(2475,1100){\makebox(0,0){\strut{} 0}}%
\put(3932,1100){\makebox(0,0){\strut{} 0.5}}%
\put(5389,1100){\makebox(0,0){\strut{} 1}}%
\put(6846,1100){\makebox(0,0){\strut{} 1.5}}%
\put(8304,1100){\makebox(0,0){\strut{} 2}}%
\put(9761,1100){\makebox(0,0){\strut{} 2.5}}%
\put(11218,1100){\makebox(0,0){\strut{} 3}}%
\put(12675,1100){\makebox(0,0){\strut{} 3.5}}%
\put(550,4600){\rotatebox{90}{\makebox(0,0){\strut{}$Z_{M_{2}}^{[1]}$}}}%
\put(7575,275){\makebox(0,0){\strut{}$M_{1}^{[0]} = \log(1 + m_{0})$}}%
\end{picture}%
\endgroup
 